\documentclass{llncs}
\usepackage{graphicx}
\usepackage{amssymb,amsmath,color}
\usepackage{caption}
\usepackage{adjustbox}
\usepackage[bookmarks=false]{hyperref}
\usepackage[anythingbreaks]{breakurl}
\usepackage[round]{natbib}
\usepackage{algorithm}
\usepackage{booktabs}
\usepackage{longtable}
\usepackage{tabu}
\usepackage{algorithmicx}
\usepackage{soul}
\usepackage{xspace}

\newcommand{\beq}{\begin{equation}}
\newcommand{\eeq}{\end{equation}}
\newcommand{\ben}{\begin{enumerate}}
\newcommand{\een}{\end{enumerate}}

\newcommand{\mc}[1]{\ensuremath{\mathcal{#1}}}
\newcommand{\bb}[1]{\ensuremath{\mathbb{#1}}}

\newcommand{\PP}{\bb{P}}

\newcommand{\Bern}{\ensuremath{\mbox{Bernoulli}}}

\newcommand{\comment}[1]{}

\renewcommand{\paragraph}[1]{\vspace{5pt}\noindent\textbf{#1\quad}}
\begin{document}
%\pagenumbering{gobble}
\pagestyle{plain}
\thispagestyle{empty}
\urlstyle{rm}

\title{Bernoulli Ballot Polling: A Manifest Improvement for Risk-Limiting Audits}

\author{
   Kellie Ottoboni\inst{1} \and
   Matthew Bernhard\inst{2} \and
   J.~Alex Halderman\inst{2} \and
   Ronald L.~Rivest\inst{3} \and
   Philip B.~Stark\inst{1} 
}
\institute{
Department of Statistics, University of California, Berkeley\and
Department of Computer Science and Engineering, University of Michigan \and
CSAIL, Massachusetts Institute of Technology}

\maketitle

\begin{abstract}
    We present a method and software for ballot-polling risk-limiting audits (RLAs) based on Bernoulli sampling:
    ballots are included in the sample with probability $p$, independently.
%    To test whether the reported outcome is correct, we use the conditional hypergeometric 
%    distribution of the number of votes for one candidate among votes for pairs of candidates.  
    Bernoulli sampling has several advantages:
    (1)~it does not require a ballot manifest; 
    (2)~it can be conducted
    independently at different locations, rather than requiring a central authority to
    select the sample from the whole population of cast ballots or requiring stratified
    sampling;
    (3)~it can start in polling places on election night, before margins are known.
%    Auditing in the polling place in turn has a number of advantages:
%    (a)~It distributes the workload. 
%    (b)~It reduces the risk that the paper trail has been compromised before the audit.
    %    (c)~It decouples the sampling from political geography, eliminating the need for stratification.
    If the reported margins for the 2016 U.S.~Presidential election are
    correct, a Bernoulli ballot-polling audit with a risk limit of 5\% and a
    sampling rate of $p_0=1\%$ would have had at
    least a 99\% probability of confirming the outcome in 42~states. 
    (The other states were more likely to have needed
    to examine additional ballots.) 
    Logistical and security advantages that auditing in the polling
    place affords may outweigh the cost of examining more ballots than some
    other methods might require.
\end{abstract}

%\noindent
%Keywords: election integrity, nonparametric hypothesis tests, Bernoulli sampling,
%Wald's sequential probability ratio test, conditional hypothesis tests, sampling
%without replacement

\section{Introduction}
No method for counting votes is perfect, and methods that rely on computers are
particularly fragile: errors, bugs, and deliberate attacks can alter results.  The
vulnerability of electronic voting was confirmed in two major
state-funded studies, California's Top-to-Bottom Review \citep{TTBR07} and
Ohio's EVEREST study \citep{everest07}.  
More recently, at the 2017 and 2018
DEFCON hacking conferences, attendees with little or no knowledge of election
systems were able to penetrate a wide range of U.S. voting
machines~\citep{blaze17,blaze18}.
%There are countless examples of small and large failures of voting systems
%currently in use in the U.S., some of which altered electoral outcomes.
%\SUGGEST{(RLR) Do we have a good citation for this last sentence? I would be
%tempted to use Gumbel's books (``Steal This Vote'' and ``Down for the
%Count'').}
Given that Russia interfered with the 2016 U.S.~Presidential election through
an ``unprecedented coordinated cyber campaign against state election
infrastructure''~\citep{ssci18}, national security demands we protect our
elections from nation states and other advanced persistent threats.

Risk-limiting audits (RLAs) were introduced in 2007~\citep{stark08a} as a mechanism
for detecting and correcting outcome-changing errors in vote tabulation,
whatever their cause---including hacking, misconfiguration, and human error. 
RLAs have been tested in practice in California, Colorado, Indiana, Virginia, Ohio, and Denmark.  
Colorado
started conducting routine statewide RLAs in 2017~\citep{lindeman18}, and Rhode
Island passed a law in 2017 requiring routine statewide RLAs starting in 2020
(RI Gen L \S\ 17-19-37.4).
%California law requires RLAs under some circumstances~\citep{TK}, but the
%state does not yet require them routinely.
RLA legislation is under consideration in a number of other states, and
bills to require RLAs have been introduced in Congress.

In this paper, we present an RLA method based on \emph{Bernoulli random
sampling}.  With simple random sampling, the number of ballots to sample is
fixed; with Bernoulli sampling, the \emph{expected sampling rate} is fixed but the
sample size is not.  Conceptually, \emph{Bernoulli ballot polling} (BBP) decides whether to
include the $j$th ballot in the sample by tossing a biased coin that has
probability $p$ of landing heads.  The ballot is included if and only if the
coin lands heads.  
Coin tosses for different ballots are independent, but have the
same chance of landing heads.
(Rather than toss a coin for each ballot, it 
more efficient to implement Bernoulli sampling in practice using \emph{geometric skipping},
described in Section~\ref{sec:geometric-skipping}.)

The logistical simplicity of Bernoulli sampling may make it useful for election
audits.  Like all RLAs, BBP RLAs require a voter-verifiable paper record.  Like
other ballot-polling RLAs \citep{lindemanEtal12,lindemanStark12}, BBP makes no
other technical demands on the voting system.  It requires no special
equipment, and only a minimal amount of software to select and analyze the
sample---in principle, it could be carried out with dice and a pencil and
paper.  In contrast to extant ballot-polling RLAs, BBP does \emph{not} require
a ballot manifest (although it does require knowing where all the ballots are,
and access to the ballots).  BBP is inherently local and parallelizable,
because the decision of whether to include any particular ballot in the sample
does not depend on which other ballots are selected, nor on how many other
ballots have been selected, nor even on how many ballots were cast.  We shall
see that this has practical advantages\@.

Bernoulli sampling is well-known in the survey sampling literature, but it is
used less often than simple random sampling, for a number of reasons.  The
variance of estimates based on Bernoulli samples tends to be larger than for
simple random samples \citep{sarndal03}, due to the fact that both the sample
and the sample size are random.  Moreover, this added randomness complicates
rigorous inferences.  A common estimator of the population mean from a
Bernoulli sample is the Horvitz-Thompson estimator, which has a high variance
when the sampling rate $p$ is small.  Often, $P$-values and confidence
intervals for the Horvitz-Thompson estimator are approximated using the normal
distribution \citep{lohr09, cochran77, thompson97}, which may be quite
inaccurate if the population distribution is skewed---as it often is in
auditing problems~\citep{panel88}. 

Instead of relying on parametric approximations, we develop a test based on
Wald's sequential probability ratio test \citep{wald45}.  The test is akin to
that in extant ballot polling RLA methods
\citep{lindemanEtal12,lindemanStark12}, but the mathematics are modified to
work with Bernoulli random samples, including the fact that Bernoulli samples
are samples without replacement.  (Previous ballot-polling RLAs relied on
sampling with replacement.) Conditional on the attained sample size $n$, a
Bernoulli sample of ballots is a simple random sample.  We maximize the
conditional $P$-value of the null hypothesis (that the reported winner did not
win) over a nuisance parameter, the total number of ballots with valid votes 
for either of a given pair of candidates, excluding invalid ballots or ballots for other candidates.  A
martingale argument shows that the resulting test is sequential: if the test
does not reject, the sample can be expanded using additional rounds of
Bernoulli sampling (with the same or different expected sampling rates) and the
resulting $P$-values will still be conservative.

A BBP RLA can begin in polling places on election night.
Given an initial sampling rate to be used across all precincts and vote centers,
poll workers in each location determine which ballots will be examined in the audit,
independently from each other and independently across ballots, 
and record the votes cast on each ballot selected. 
(Vote-by-mail and provisional ballots can be audited similarly; 
see Section~\ref{sec:vbm-provisional}.)
Once the election results are reported, the sequential probability ratio test can be applied
to the sample results to determine whether there is sufficient evidence that the reported
outcome is correct.\footnote{%
A variant of the method does not require the reported results (the current method
used the reported results to construct the alternative hypothesis). 
We do not present that method here; it is related to ClipAudit \citep{rivest17}.
}
If the sample does not provide sufficiently strong evidence to attain the risk limit, 
the sample can be expanded using subsequent rounds of Bernoulli sample until either 
the risk limit is attained or all ballots are inspected.
Figure~\ref{fig:recommended-audit} summarizes the procedure.

%% \TODO{What happened to the rule of thumb?}
%% We provide a rule of thumb for choosing the sampling rate
%% based on the expected margin and number of ballots cast in the contest.  The
%% sampling rate decreases quadratically in the vote margin; e.g., the initial
%% sampling rate goes down by a factor of four if the anticipated margin is doubled.

BBP has a number of practical advantages, with little additional workload
in terms of the number of ballots examined.
Workload simulations show that the number of ballots needed to confirm a correctly
reported outcome is similar for BBP and the BRAVO RLA \citep{lindemanEtal12}.
If the choice of initial sampling rate (and thus, the initial sample size) is larger than necessary, 
the added efficiency of conducting the audit ``in parallel'' across the entire election may 
outweigh the cost of examining extra ballots.
Using statewide results from the 2016 United States presidential election,
BBP with a 1\% initial sampling rate would have had at least a 99\% chance 
of confirming the results in 42 states (assuming the reported results were in fact correct).
A Python implementation of BBP is
available at [\emph{omitted for blind review}].

\begin{figure*}[t]
    \centering
%\floatpagestyle{empty}    % suppresses page number on this float page
%\scalebox{.8}{
\framebox[\textwidth]{
  \parbox{0.95\textwidth}{
    \vspace{8pt}
    \centerline{\bf\emph{\Large Procedure for a Bernoulli ballot-polling audit}}
    \vspace{-3pt}
\begin{enumerate}
\item {\bf Set initial sampling rate.} 
  Choose initial sampling rate $p_0$, based on pre-election polls or set at a fixed value.
  If $p_0$ is selected based on an estimated margin, use the ASN heuristic
  in Section~\ref{sec:power}.
\item {\bf Sample ballots and record audit data.}
  Use geometric skipping (below) with rate $p_0$ to select ballots to inspect.
  Record votes on all inspected ballots.
\item {\bf Check attained risk.} 
  Once the final election results have been reported, for each contest under audit and
  for each reported $(\text{winner},\text{loser})$ pair $(w, \ell)$:
  \begin{itemize}
  \item Calculate $B_w$, $B_\ell$, and $B_u$ from the audit sample.
  \item Find the (maximal) $P$-value from $B_w, B_\ell, B_u$.
  \end{itemize}
\item{\bf Escalate if necessary.}  If, for any $(w, \ell)$ pair, the $P$-value is greater than $\alpha$, expand the audit in one of the ways 
described in Section~\ref{sec:escalation}.
\end{enumerate}
\hrule\medskip
\centerline{\bf\emph{\large Procedure for geometric skip sampling}}
\vspace{-6pt}
\begin{enumerate}
\item {\bf Set the random seed.} 
  In each polling place, use a cryptographically secure
        PRNG, such as SHA-256, with a seed chosen using true randomness.
        \item {\bf Sample ballots} following Section~\ref{sec:geometric-skipping}.
  For each bundle of ballots: Set $Y_0 = 0$ and set $j = 0$.
  \begin{itemize}
  \item $j \leftarrow j+1$
  \item Generate a uniform random variable $U$ on $[0, 1)$.
  \item $Y_j \leftarrow \left\lceil \frac{\ln(U)}{\ln(1 - p)} \right\rceil$.
  \item If $\sum_{k=1}^j Y_j$ is greater than the number of ballots in the bundle, stop. Otherwise, skip the next $Y_j-1$ ballots in the bundle, and include the ballot after that one (i.e., include ballot $\sum_{k=1}^j Y_j$)
  \end{itemize}
\end{enumerate}
}
}
%}
\normalsize
\caption{\bf Bernoulli ballot-polling audit step-by-step procedures.}
\label{fig:recommended-audit}
\end{figure*}

\section{Notation and Mathematical Background}

We consider social choice functions that are variants of majority and plurality
voting: the winners are the $k \ge 1$ candidates who receive the most votes.
This includes ordinary ``first-past-the-post'' contests, as well as ``vote for
$k$'' contests.\footnote{%
  The same general approach works for some preferential voting schemes, such as
  Borda count and range voting, and for proportional representation schemes such
  as D'Hondt~\citep{starkTeague14}.
  We do not consider instant-runoff voting (IRV).
}
As explained in \cite{lindemanEtal12}, it suffices to consider one $(\text{winner}, \text{loser})$ pair at a time:
the contest outcome is correct if every reported winner actually received more votes than
every reported loser.
Auditing majority and super-majority contests requires only minor modifications.\footnote{%
For instance, for a majority contest, one simply pools the votes for all the reported losers into
a single ``pseudo-candidate'' who reportedly lost.
}
Section~\ref{sec:multipleContests} addresses auditing multiple contests simultaneously.
%\footnote{%
%  Generally, the extension is simple because the multiplicity of tests is in the conservative direction:
%  there is a full hand count unless \emph{all} the null hypotheses are rejected.
%}

Let $w$ denote a reported winning candidate and $\ell$ denote a reported losing
candidate.  Suppose that the population contains $N_w$ ballots with a valid
vote for $w$ but not $\ell$, $N_\ell$ ballots with a valid vote for $\ell$ but
not $w$, and $N_u$ ballots with votes for both $w$ and $\ell$ or for neither
$w$ nor $\ell$.  The total number of ballots is $N = N_w + N_\ell + N_u$.  Let
$N_{w\ell} \equiv N_w + N_\ell$ be the number of ballots in the population with
a valid vote for $w$ or $\ell$ but not both.  For Bernoulli sampling, $N$ may
be unknown; in any event, $N_w, N_\ell$, and $N_u$ are unknown, or the audit
would not be necessary.

If we can reject the null hypothesis that $N_\ell \ge N_w$ at significance
level $\alpha$, we have statistically confirmed that $w$ got more votes than
$\ell$.  Section~\ref{sec:tests} presents a test for this hypothesis that
accounts for the nuisance parameter $N_{w\ell}$.  We assume that ties are
settled in a deterministic way and that if the audit is unable to confirm the
contest outcome, a full manual tally resulting in a tie would be settled
in the same deterministic way.

\subsection{Multi-round Bernoulli Sampling} \label{sec:multi-round}
A \emph{$\Bern(p)$ random variable} $\mc{I}$ is a random variable that takes
the value~$1$ with probability $p$ and the value $0$ with probability $1-p$.
BBP uses Bernoulli sampling, which involves independent selection of different
ballots with the same probability $p$ of selecting each ballot: $\mc{I}_j = 1$ if
and only if ballot $j$ is selected to be in the sample, where $\{ \mc{I}_j
\}_{j=1}^N$ are independent, identically distributed (IID) $\Bern(p)$ random
variables.

Suppose that after tossing a $p_0$-coin for every item in the population,
we toss a $p_1$-coin for every item (again, independently), 
and include an item in the sample if the first or 
second toss for that item landed heads. 
That amounts to drawing a Bernoulli sample using
selection probability $1-(1-p_0)(1-p_1)$: an item is in the sample unless its coin
landed tails on both tosses, which has probability $(1-p_0)(1-p_1)$.  This extends to
making any integral number $K$ of passes through the population of ballots, 
with pass $k$ using a coin that has chance $p_k$ of landing heads: such ``$K$-round''
Bernoulli sampling is still Bernoulli sampling, with $\PP \{ \mc{I} = 1 \} = p = 1
- \prod_{k=0}^{K-1}(1-p_k)$.

\subsection{Exchangeability and Conditional Simple Random Sampling}
Because the $N$ variables $\{\mc{I}_j\}$ are IID, they are \emph{exchangeable},
meaning their joint distribution is invariant under the action of the symmetric group (relabelings).
Consider a collection of indices $\mc{S} \subset \{1, \ldots, N\}$ of size $k$, $0 \le k \le N$.
Define the event
\[
   \mc{I}_{\mc{S}} \equiv \{ \mc{I}_j = 1, \forall j \in \mc{S}, \mbox{ and } \mc{I}_j = 0, \forall j \notin \mc{S} \}.
\]
Because $\{\mc{I}_j\}$ are exchangeable, $\PP \mc{I}_{\mc{S}} = 
\PP \mc{I}_{\mc{T}}$ for every set $\mc{T} \subset \{1, \ldots, N\}$ of size $k$,
since every such set $\mc{T}$ can be mapped to $\mc{S}$ by a one-to-one relabeling of the indices.

It follows that, conditional on the attained size of the sample, $n = \sum_{j=1}^N \mc{I}_j$,
all ${N \choose n}$ subsets of size $n$ drawn from the $N$ items are equally likely: 
the sample is conditionally a simple random sample (SRS) of size $n$.
This is foundational for the methods we develop. 

\section{Tests}
\label{sec:tests}

Suppose we draw a Bernoulli sample of ballots.
The random variable $B$ is the number of ballots in the sample.
Let $B_w$ denote the number of ballots in the sample with a vote for $w$ but not $\ell$;
let $B_\ell$ denote the number of ballots in the sample with a vote for $\ell$ but not $w$;
and let $B_u$ denote the number of ballots in the sample with a vote for both $w$ and $\ell$
or neither $w$ nor $\ell$, so $B = B_w + B_\ell + B_u$.

\subsection{Wald's SPRT with a Nuisance Parameter}

We want to test the compound hypothesis that $N_w \le N_\ell$ against the alternative that 
$N_w = V_w$, $N_\ell = V_\ell$, and $N_u = V_u$, with $V_w - V_\ell > 0$.\footnote{%
The alternative hypothesis is that the reported results are correct;
as mentioned above, there are other alternatives one could use that do not depend on the reported results, but
we do not present them here.
}
We present a test based on Wald's sequential probability ratio test (SPRT) \citep{wald45}.

The values $V_w$, $V_\ell$, and $V_u$ are the reported results (or values related to those reported results; see \cite{lindemanEtal12}). 
In this problem, $N_u$ (equivalently, $N_{w\ell} \equiv N_w + N_\ell$) is a nuisance parameter: we care about $ N_w - N_\ell$, the margin of the reported winner over the reported loser.

Conditional on $B=n$, the sample is a simple random sample.
The conditional probability that the sample will yield counts $(B_w, B_\ell, B_u)$ under the alternative hypothesis is
$$
    \frac{\prod_{i=0}^{B_w-1} (V_w-i) \; 
             \prod_{i=0}^{B_\ell-1} (V_\ell-i) \;
             \prod_{i=0}^{B_u-1} (V_u-i)}
            {\prod_{i=0}^{n-1} (N-i)}.
$$
If $B_\ell \ge B_w$, the data obviously do not provide evidence against the null, so we suppose that
$B_\ell < B_w$, in which case, the element of the null that will maximize the probability of the observed data has $N_w = N_\ell$.
Under the null hypothesis, the conditional probability of observing $(B_w, B_\ell, B_u)$ is
$$
    \frac{ \prod_{i=0}^{B_w-1} (N_w-i) \;
             \prod_{i=0}^{B_\ell-1}(N_w - i)
             \prod_{i=0}^{B_u-1} (N_u-i)}
             {\prod_{i=0}^n (N-i)},
$$
for some value $N_w$ and the corresponding $N_u = N - 2N_w$.
How large can that probability be if the null hypothesis is true? 
The probability under the null is maximized by any integer 
$x \in \{ \max(B_w, B_\ell), \ldots, (N-B_u)/2 \}$ that maximizes 
$$
\prod_{i=0}^{B_w-1} (x-i) \; \prod_{i=0}^{B_\ell-1} (x-i) \; \prod_{i=0}^{B_u-1} (N - 2x - i).
$$
The logarithm is monotonic, so any maximizer $x^*$ also maximizes
$$ f(x) = \sum_{i=0}^{B_w-1} \ln (x-i) + \sum_{i=0}^{B_\ell-1} \ln (x-i) + \sum_{i=0}^{B_u-1} \ln (N-2x - i).$$

The second derivative of $f$ is everywhere negative, so $f$ is convex and has a unique 
real-valued maximizer on $[ \max(B_w, B_\ell), (N - B_u)/2]$,
either at one of the endpoints or somewhere in the interval.
The derivative $f'(x)$ is
$$ f'(x) = \sum_{i=0}^{B_w-1} \frac{1}{ x-i } + \sum_{i=0}^{B_\ell-1} \frac{1}{ x-i} -2 \sum_{i=0}^{B_u-1} \frac{1}{N-2x - i}.$$

If $f'(x)$ does not change signs, then the maximum is at one of the endpoints,
in which case $x^*$  is the endpoint for which $f$ is larger.
Otherwise, the real maximizer occurs at a stationary point.
If the real-valued maximizer is not an integer, convexity guarantees that the integer maximizer $x^*$ is one
of the two integer values that bracket the real maximizer: either  $\lfloor x \rfloor$ or $\lceil x \rceil$.

%The first two terms on the right increase monotonically with $x$ and the last term decreases monotonically with $x$.
%This yields bounds without having to evaluate $f$ everywhere.
%Suppose $y < z$. 
%Then for all integer $x$ between $y$ and $z$, 
%$$ f(x) \le \sum_{i=0}^{W_n-1} \ln (z-i) + \sum_{i=0}^{L_n-1} \ln (z-c-i) + \sum_{i=0}^{U_n-1} \ln (N_s-2y+c-i).$$
%The optimization problem can be solved using a branch and bound approach.
%For instance, start by evaluating 
%$$
%   f^+(x) \equiv \sum_{i=0}^{W_n-1} \ln (x-i) + \sum_{i=0}^{L_n-1} \ln (x-c-i)
%$$
%and
%$$
%  f^-(x) \equiv \sum_{i=0}^{U_n-1} \ln (N_s-2x+c-i)
%$$
%at $\max(W_n, L_n+c)$, $(N_s-U_n)/2$, and their midpoint,
% to get the values of $f = f^+ + f^-$ at those three points, along with 
%upper bounds on $f$ on the ranges between them.
%At stage $j$, we have evaluated $f$, $f^+$, and $f^-$ at $j$ points $x_1 < x_2 < \ldots < x_j$, and 
%we have upper bounds on $f$ on the $j-1$ ranges $R_m = \{x_m, x_m+1, \ldots, x_{m+1}\}$
%between those points.
%Let $U_m$ be the upper bound on $f(x)$ for $x \in R_m$.
%Suppose that for some $h$, $f(x_h) = \max_{m=1}^j U_m$.
%Then $x^* = x_h$ is a global maximizer of $f$.
%If there is some $U_m > \max_i f(x_i)$, then subdivide the range with the largest $U_m$,
%calculate $f$, $f^+$, and $f^-$ at the new point, and repeat.
%This algorithm must terminate by identifying a global maximizer $x^*$ after a finite number of steps.

A conservative $P$-value for the null hypothesis after $n$ items have been drawn is thus
$$
   P_n =  \frac{\prod_{i=0}^{B_w-1} (x^*-i) \; \prod_{i=0}^{B_\ell-1}  (x^*-i) \; \prod_{i=0}^{B_u-1} (N-2x^*-i)}{\prod_{i=0}^{B_w-1}(V_w-i) \; \prod_{i=0}^{B_\ell-1} (V_\ell-i) \; \prod_{i=0}^{B_u-1} (V_u-i)}.
$$

The SPRT is appealing because it leads to an elegant escalation method if the first round of
Bernoulli sampling does not attain the risk limit: simply make another round of Bernoulli sampling,
as described in Section~\ref{sec:escalation}.
If the null hypothesis is true, then $\Pr \{ \inf_k P_k < \alpha \} \le \alpha$, where
$k$ counts the rounds of Bernoulli sampling.
That is, the risk limit remains conservative for any number of rounds of Bernoulli sampling.

%
%\subsection{Conditional Hypothesis Testing}
%\label{sec:conditional}
%The tests we derive condition on the attained size $m$ of the Bernoulli 
%sample drawn from some subset of the
%$N$ ballots (for instance, the subset of ballots that 
%show either a valid vote for $w$ or for $\ell$ but not both).
%If the conditional tests are always conducted at significance level $\alpha$ or less, i.e., so that
%$\PP \{\mbox{Type I error} | n = m\} \le \alpha$, then the
%overall procedure has significance level $\alpha$ or less:
%\begin{eqnarray}
%    \PP \{\mbox{Type I error}\} &=& \sum_{m=0}^N \PP \{\mbox{Type I error} | n = m\} \PP \{n=m \} \nonumber \\
%       & \le & \sum_{m=0}^N \alpha \PP \{ n=m \}  =  \alpha.
%\end{eqnarray}

\subsection{Auditing Multiple Contests}
\label{sec:multipleContests} 
The math extends to audits of multiple contests; we omit the derivation, but 
see, e.g., \citet{lindemanStark12}.
The same sample can be used to audit any number of contests simultaneously.
The audit proceeds to a full hand count unless every null hypothesis is
rejected, that is, unless we conclude that \emph{every} winner beat
\emph{every} loser in \emph{every} audited contest.  
The chance of rejecting
all those null hypotheses cannot be larger than the smallest chance of
rejecting any of the individual hypotheses, because the probability of an
intersection of events cannot be larger than the probability of any one of
the events.  The chance of rejecting any individual null hypothesis is at
most the risk limit, $\alpha$, if that hypothesis is true.  Therefore the
chance of the intersection is not larger than $\alpha$ if any contest outcome
is incorrect: the overall risk limit is $\alpha$, with no need to adjust for multiplicity.

%There are $N$ ballots in all, of which $N\pi_i$ are for candidate $i$, $i = 1,
%\ldots, C$.  It is claimed that the candidates with indices in $\mc{W}$ are
%winners and those with indices in $\mc{L}$ are losers, where $\mc{W} \cup
%\mc{L} = \{1, \ldots, C\}$.  Because the social choice function is plurality,
%this amounts to the claim that $\pi_w > \pi_\ell$ for every $(w, \ell)$ pair,
%where $w \in \mc{W}$ is a reported winner and $\ell \in \mc{L}$ is a reported
%loser.  If we can reject the hypothesis $\pi_\ell \ge \pi_w$ for every $(w,
%\ell)$ pair at significance level $\alpha$, we have statistically confirmed the
%electoral outcome.

\section{Escalation} \label{sec:escalation}
If the first round of Bernoulli sampling with rate $p_0$ does not generate
strong evidence that the election outcome is correct, we have several options:

\begin{enumerate}
   \item conduct a full hand count
   \item augment the sample with additional ballots selected in some manner, for instance, 
             making additional rounds of Bernoulli sampling, possibly with different values of $p$
   \item draw a new sample and use a different auditing method, \emph{e.g.}, ballot-level comparison auditing
\end{enumerate}

The first approach is always conservative. Both the second and third
approaches require some statistical care, as repeated testing introduces additional
opportunities to wrongly conclude that an incorrect election outcome is
correct. 

To make additional rounds of Bernoulli sampling, it may help to keep track of which ballots
have been inspected.\footnote{%
Once ballots are aggregated in a
precinct or scanned centrally, it is unlikely that they will stay in the
same order. 
} 
That might involve stamping audited ballots with
``audited'' in red ink, for example. 

Section~\ref{sec:multi-round} shows that
if we make an integral number of passes through the
population of ballots, tossing a $p_k$-coin for each as-yet-unselected item (we only toss
the coin for an item on the $k$th pass if the coin has not landed heads for
that item in any previous pass), then the resulting sample is
a Bernoulli random sample with selection probability $p = 1-\prod_{k=0}^{K-1} p_k$.  
Conditional on the sample size $n$ attained after $K$ passes, 
every subset of size $n$ is equally likely to be selected.
Hence, conditional on the event that $n$ tosses gave heads, the sample is a simple 
random sample of size $n$ from the $N$ ballots.

The SPRT applied to multi-round Bernoulli sampling is conservative: the unconditional chance of
rejecting the null hypothesis if it is true is at most $\alpha$, because, if
the null is true, the chance that the SPRT exceeds $1/\alpha$ for \emph{any}
$K$ is at most $\alpha$.

The third approach allows us to follow BBP with a different, more efficient approach,
such as ballot-level comparison auditing \citep{lindemanStark12}.
%(for instance, if the ballots were stored carefully and tracked in a ballot
%manifest, a standard ballot-polling RLA could be done), 
This may require steps to ensure that multiplicity does not make the
risk larger than the nominal risk limit, e.g., by adjusting the risk limit using Bonferroni's inequality.
%One way to control the overall chance of confirming an incorrect outcome when
%the audit has more than one sampling round is through the crude Bonferroni
%bound used by \citet{stark08a}, which partitions the allowable risk $\alpha$
%into pieces allocated to different sampling rounds.  For instance, the first
%round might use a risk limit of $2\alpha/3$, the second a risk limit of
%$\alpha/3$, and if the second round does not confirm the outcome, a full hand
%count is conducted.
%
%More generally, suppose the audit is allowed to have $A$ stages before calling
%for a full manual tally.  Then if the audit at stage $a$ is conducted to a risk
%limit of $\alpha_a$, and $\sum_{a=1}^A \alpha_a \le \alpha$, the overall risk
%limit is no larger than $\alpha$.  The escalation procedure enabled by using
%the sequential test is more elegant, and perhaps less conservative, than this
%approach.

\section{Initial Sampling Rate}
\label{sec:power}

We would like to choose the initial sampling rate $p_0$ sufficiently large
that a test of the hypothesis $N_w \le N_\ell$ will have high power against the
alternative $N_w = V_w, N_\ell = V_\ell$, with $V_w - V_\ell = c$ for modest
margins $c>0$, but not so large that we waste effort.

%A margin of 1\% between candidates $w$ and $\ell$ corresponds to $\mu =
%0.005$.
%
%Under the null, the distribution of $n_w$ given $N_{w\ell}$ is
%\Hypergeom$(N_{w\ell}/2, N_{w\ell}, n_{w\ell})$.  $N_w = \frac{N_{w\ell}}{2}$.
%The threshold at which we reject the null hypothesis is the $1-\alpha$
%quantile of this distribution.  Let $q_\alpha$ denote this quantile, so that
%$\mathbb{P}(n_w \geq q_\alpha || N_w = \frac{N_{w\ell}}{2}) \leq \alpha$. 
%
%The power of the test against the point alternative $N_w =
%\frac{N_{w\ell}}{2}+\epsilon$ with $\epsilon>0$, conditional on having drawn
%$B_{w\ell} = n_{w\ell}$, is $\mathbb{P}(n_w \geq q_\alpha || N_w =
%\frac{N_{w\ell}}{2}+\epsilon)$.
%
%In Bernoulli sampling, $B$ is not fixed; it has a binomial distribution with
%parameters $N$ and $p$.  The unconditional power of the test is therefore:
%\begin{align} \mathbb{P}(n_w \geq q_\alpha || N_w =
%\frac{N_{w\ell}}{2}+\epsilon) &= \sum_{n=0}^{N_{w\ell}}
%\mathbb{P}(\text{sample } n \text{ out of } N_{w\ell})\mathbb{P}(n_w \geq
%q_\alpha ||n, N_w = \frac{N_{w\ell}}{2}+\epsilon) \nonumber \\
%%&= \sum_{n=0}^{N_{w\ell}} {N_{w\ell} \choose n} p^{n}(1-p)^{N_{w\ell}-n} \mathbb{P}(n_w \geq q_\alpha || n, N_w = \frac{N_{w\ell}}{2}+\epsilon) \nonumber \\
%&= \sum_{n=0}^{N_{w\ell}}\sum_{j=q_\alpha}^n {N_{w\ell} \choose n}
%p^{n}(1-p)^{N_{w\ell}-n} \frac{{ {\frac{N_{w\ell}}{2}+\epsilon} \choose j }{
%{\frac{N_{w\ell}}{2}-\epsilon} \choose n-j }}{{N_{w\ell}\choose n}}.
%\end{align}

There is no analytical formula for the power of the sequential hypothesis test under this sampling
procedure, but we can use simulation to estimate the sampling rates needed to have a high 
probability of confirming correctly reported election results.
Table~\ref{tab:minmargin} gives the sampling rate $p_0$ needed to attain 80\%,
90\%, and 99\% power for a 2-candidate race in which there are no undervotes or
invalid votes, for a 5\% risk limit and a variety of margins and contest sizes.
The simulations assume that the reported vote totals are correct.
The required $p_0$ may be prohibitively large for small races
and tight margins; Section~\ref{sec:evaluation} shows that with high probability, even a 1\%
sampling rate would be sufficient to confirm the outcomes of the vast majority
of U.S.~federal races without further escalation.

The sequential probability ratio test in Section~\ref{sec:tests} is similar to the BRAVO RLA
presented in \cite{lindemanStark12} when the sampling rate is small relative to the
population size.
One difference is that BBP incorporates information about
the number of undervotes, invalid votes, or votes for candidates other than $w$ and $\ell$,
and that Bernoulli sampling is without replacement; BRAVO is based on sampling with replacement.
If every ballot has a valid vote either for $w$ or for $\ell$ and the sampling rate is small
relative to the population size, the expected workload of these two procedures is similar.
The \emph{average sample number} (ASN) \citep{wald45},
the expected number of draws required either to accept or to reject the null hypothesis,
for BRAVO using a risk limit $\alpha$ and margin $m$ is approximately
$$\text{ASN} \approx \frac{2 \ln(1/\alpha)}{m^2}.$$
This formula is valid when the sampling rate is low and the actual margin is not substantially
smaller than the (reported) margin used as the alternative hypothesis.

The ASN gives a rule of thumb for choosing the initial sampling rate.
For a risk limit of 5\% and a margin of $5\%$, the ASN is about 2,400 ballots.
For a margin of 10\%, the ASN is about 600 ballots.
These values are lower than the sample sizes implied by Table~\ref{tab:minmargin}:
the sampling rates in the table have a higher probability that the initial sample will
be sufficient to conclude the audit, while a sampling rate based on the ASN
will suffice a bit more than half of the time.\footnote{%
The distribution of the sample size is skewed to the right: the expected sample size is generally larger than the median sample size.
}
The ASN multiplied by 2--4 is a rough approximation to initial sample size needed
to have roughly a 90\% chance that the audit can stop without additional sampling, if the reported
results are correct.

The value of $p_0$ should be adjusted to account for ballots that have votes for neither 
$w$ nor $\ell$ (or for both $w$ and $\ell$).  
If $r=\frac{N_u}{N}$ is the fraction of such ballots,
the initial sampling rate $p_0$ should be inflated by a factor of $\frac{1}{1-r}$.  
For example, if half of the
ballots were undervotes or invalid votes, then the sampling rate would need to
be doubled to achieve the same power as if all of the ballots were valid votes
for either $w$ or $\ell$.

\begin{table}[t]
\centering
\caption{\protect \label{tab:minmargin}
{\bf Estimated sampling rates needed for Bernoulli ballot polling}
for a 2-candidate race with a $5\%$ risk limit.
These simulations assume the reported margins were correct.
%Source code is available at (\emph{blinded for review}).
%\url{www.github.com/research/???}.
}
\vspace{0.75em}
\begin{tabu} to \linewidth {rX[1.25r]X[0.25cm]X[r]X[r]X[r]}
\toprule  
          &                & & \multicolumn{3}{c}{sampling rate $p$ to achieve $\ldots$} \cr
\cmidrule{4-6}
true margin & ballots cast & & 80\% power & 90\% power & 99\% power \cr
\midrule
   1\%    &    100,000     & & 55\% & 62\% & 77\% \cr
   2\%    &    100,000     & & 23\% & 30\% & 46\% \cr
   5\%    &    100,000     & &  5\% &  7\% & 12\% \cr
  10\%    &    100,000     & &  2\% &  2\% &  4\% \cr
  20\%    &    100,000     & &  1\% &  1\% &  1\% \cr
\cmidrule{1-6}
   1\%    &  1,000,000     & &  10.4\% &  14.2\% & 24.2\% \cr
   2\%    &  1,000,000     & &  2.9\% &  4.0\% &  7.5\% \cr
   5\%    &  1,000,000     & &  0.5\% &  0.7\% &  1.3\% \cr
  10\%    &  1,000,000     & &  0.2\% &  0.2\% &  0.4\% \cr
  20\%    &  1,000,000     & &  0.1\% &  0.1\% &  0.1\% \cr
\cmidrule{1-6}
   1\%    &  10,000,000     & &  1.15\% &  1.66\% & 3.11\% \cr
   2\%    &  10,000,000     & &  0.30\% &  0.42\% &  0.84\% \cr
   5\%    &  10,000,000     & &  0.05\% &  0.07\% &  0.13\% \cr
  10\%    &  10,000,000     & &  0.02\% &  0.02\% &  0.04\% \cr
  20\%    &  10,000,000     & &  0.01\% &  0.01\% &  0.01\% \cr
\bottomrule
\end{tabu}
\end{table}

\section{Implementation}

\subsection{Election Night Auditing}
\label{sec:electionday}

Previous approaches to auditing require a sampling frame (possibly stratified, \emph{e.g.}, by
mode of voting or county).
That requires knowing how many ballots there are in each stratum.
In contrast, Bernoulli sampling makes it possible to start the audit
at polling places immediately after the last vote has been cast in that polling place,
without even having to count the ballots cast in the polling place.
This has several advantages:
\begin{enumerate}
   \item It parallelizes the auditing task and can take advantage of staff (and observers) who are already on site at polling places.
   \item It takes place earlier in the chain of custody of the physical ballots, before the ballots are exposed to some risks of loss, addition, substitution, or alteration.
   \item It may add confidence to election-night result reporting.
\end{enumerate}

The benefit is largest if $p_0$ is large enough to allow the audit to complete without
escalating.
Since reported margins will not be known on election night, 
$p_0$ might be based on pre-election polls, 
or set to a fixed value.  
There is, of course, a chance that the initial sample will not suffice to confirm outcomes,
either because the true margins are smaller than anticipated, or because the
election outcome is in fact incorrect.

There are reasons polling-place BBP audits might not be desirable.
\begin{enumerate}
    \item Pollworkers, election judges, and observers are likely to be tired and ready to go home when
    polls close.
    \item The training required to conduct and to observe the audit goes beyond what poll workers
    and poll watchers usually receive.
    \item Audit data need to be captured and communicated reliably to a central authority to compute
    the risk (and possibly escalate the audit) after election results are reported.
\end{enumerate}

\subsection{Vote-by-mail and Provisional Ballots}
\label{sec:vbm-provisional}
The fact that Bernoulli sampling is a ``streaming'' algorithm may help simplify logistics
compared with other sampling methods.
For instance. Bernoulli sampling can be used with vote-by-mail (VBM) ballots.
Bernoulli sampling can also be used with provisional ballots.
VBM and provisional ballots can be sampled
as they arrive (after signature verification), or aggregated, e.g., daily or weekly.  
Ballots do not need to be opened or
examined immediately in order to be included in the sample: they can be set aside and
inspected after election day or after their provisional status has been
adjudicated.  Any of
these approaches yields a Bernoulli sample of all ballots cast in the
election, provided the same value(s) of $p$ are used throughout.

\subsection{Geometric Skipping} \label{sec:geometric-skipping}
In principle, one can implement Bernoulli sampling by actually rolling dice, or 
by assigning a $U[0,1]$
random number to each ballot, independently across ballots.  A ballot is in the
sample if and only if its associated random number is less than or equal to
$p$.

However, that places an unnecessarily high burden on the quality of the
pseudorandom number generator---or on the patience of the people responsible
for selecting ballots by mechanical means, such as by rolling dice.  If the ballots are in
physical groups (e.g., all ballots cast in a precinct), it can be more efficient
to put the ballots into some canonical order (for instance, the order in which they are
bundled or stacked) and to rely on the fact that the \emph{waiting times} between 
successes in independent Bernoulli$(p)$ trials are independent Geometric$(p)$ random
variables: the chance that the next time the coin lands heads will be
$k$th tosses after the current toss is $p(1-p)^{k-1}$.

To select the sample, instead of generating a
Bernoulli random variable for every ballot, we suggest generating a sequence
of geometric random variables $Y_1, Y_2, \ldots$\@ The first ballot in the sample
is the one in position $Y_1$ in the group, the second is the one in position
$Y_1 + Y_2$, and so on.  We continue in this way until $Y_1 + \ldots + Y_j$ is
larger than the number of ballots in the group.
This \emph{geometric skipping} method is implemented in the software we provide.
%, that is, until we run out of ballots.

\subsection{Pseudorandom Number Generation}
\label{sec:prng}
%In practice, it is difficult to generate a large quantity of true random
%numbers.  Instead, one uses a \emph{pseudorandom number generator} (PRNG), a
%deterministic algorithm that outputs numbers that are indistinguishable from
%random.  Some PRNGs are better than others.  
To draw the sample, we propose using a cryptographically secure PRNG based on the SHA-256
hash function, setting the seed using 20 rolls of 10-sided dice, in a public ceremony.
This is the method that the State of Colorado uses to select the sample for risk-limiting audits.

This is a good choice for election audits for several reasons.  
First, given the initial seed, anyone can verify that the sequence of ballots audited
is correct.
Second, unless the seed is known, the ballots to be audited
are unpredictable, making it difficult for an adversary to ``game''
the audit.  
Finally, this family of PRNGs produces high-quality 
pseudorandomness. 

Implementations of SHA-256-based PRNGs are available in many languages, including
Python and Javascript.
The code we provide for geometric skipping
relies on the \texttt{cryptorandom} Python library, which implements such a
PRNG. 

While Colorado sets the seed for the entire state in a public ceremony, it may be 
more secure to generate seeds for polling-place audits locally,
after the ballots have been collated
into stacks that determine their order for the purpose of the audit.
If the seed were known before the order of the ballots was fixed, 
an adversary might be able to arrange that the 
ballots selected for auditing reflect a dishonest outcome. 

%% \SUGGEST{(RLR) We might mention that although the sequence of PRNs is verifiable, the currently
%%   proposed framework does not enable post facto verification that the ballots pulled were the correct ones;
%%   only observers present at the ballot-pulling can verify that.  This is different that calling for
%%   specific ballots that have pre-stamped numbers on them, for example.}

%To map PRNs that are uniformly distributed on the interval $[0, 1]$ into a
%Geometric($p$) random variable, one can use the inverse transform method.  If
%$U$ has a uniform distribution on $[0, 1]$, then the random variable
%$\left\lceil \frac{\ln(U)}{\ln(1 - p)} \right\rceil$ has a geometric
%distribution with parameter $p$ \citep{devroye86}.

While the sequence of ballots selected by this method is
verifiable, there is no obvious way to verify \emph{post facto} 
that the ballots examined were the correct ones.  Only
observers of the audit can verify that. 
Observers' job would be easier if ballots were pre-stamped with (known)
unique identifiers, but that might compromise vote anonymity.

\section{Evaluation}
\label{sec:evaluation}

%\subsection{Simulation}
\begin{figure}[t]
    \centering
%    \scalebox{.8}{
    \includegraphics[width=.95\columnwidth]{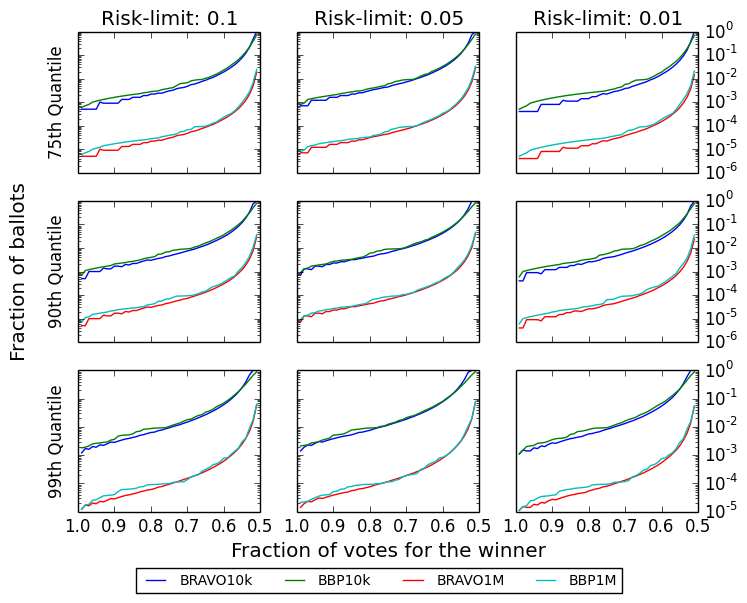}
%    }
    \caption{\textbf{Simulated quantiles of sample sizes by fraction of votes
    for the winner for a two candidate race} in elections with 10,000 ballots
    and 1~million ballots, for BRAVO ballot-polling audits (BPA) and Bernoulli
    ballot polling audits (BBP), for various risk-limits. The simulations
    assume every ballot has a valid vote for one of the two candidates. 
    % \TODO{Get rid of the method we are not writing
    % about.  Expand the caption (I can't figure out what this is showing). Label
    % the axes better. Change BPA to BRAVO. Clarify: this is a 2-candidate
    % contest with no invalid ballots? The margin is a fraction?}
    \label{fig:comp_plot}
}
\end{figure}

As discussed in Section~\ref{sec:power}, we expect that \emph{workload} (total
number of ballots examined) for Bernoulli ballot polling to be approximately
the same as BRAVO ballot polling.  Figure~\ref{fig:comp_plot} compares the
fraction of ballots examined for BRAVO audits and BBP for a 2-candidate
contest, estimated by simulation.
The simulations use contest sizes of 10,000 and 1,000,000 ballots,
each of which has either a valid vote for the winner or a valid vote for the loser.
The percentage of votes for the winner
ranges from 99\% (almost all the votes go to the winner) to 50\% (a tie). 
The methods produce similarly shaped curves; BBP requires slightly
more ballots than BRAVO.

As the workload of BRAVO and BBP are similar, the cost of running a Bernoulli
audit should be similar to BRAVO\@.  
There are likely other efficiencies to
Bernoulli audits, \emph{e.g.}, if the first stage of the audit can be completed on
election night in parallel, it might result in lower cost as election workers
and observers would not have to assemble in a different place and time for the
audit.
Even if the cost were somewhat higher, that might be offset by
advantages discussed in Section~\ref{sec:discussion}. 

\subsection{Empirical Data}

We evaluate BBP using precinct-level data from the 2016 U.S.\ presidential
election, collected from OpenElections~\citep{openelex18} or by hand where that
dataset was incomplete. If the reported margins are correct, BBP 
with a sampling rate of $p_0 = 1\%$ and a risk-limit of
5\% would have a 99\% or higher chance of confirming the outcome in 42 states.
The mean sample size per-precinct for this method is about 10 ballots,
indicating that if the audit is conducted in-precinct the workload will be
fairly minute. There is thus a large probability that if the election outcomes
in those states are correct, they would not have to audit additional ballots
beyond the initial sample.

\section{Discussion}
\label{sec:discussion}
Bernoulli ballot polling has a number of practical advantages.  We
have discussed several throughout the paper, but we review all of them
here:
\begin{itemize}
  \item It reduces the need for a ballot manifest: ballots
    can be stored in any order, and the number of ballots in a given
    container or bundle does not need to be known to draw the sample.  
  \item The work can be conducted in parallel across polling places, and can be performed by workers (and observed by members of the public) already in place on
    election day. 
  \item The same sampling method can be used for polling places, vote centers, VBM, and provisional ballots, without the need to stratify the sample explicitly.
  \item If the initial sampling rate is adequate, the winners can be confirmed
    shortly after voting finishes---perhaps even at the same
    time that results are announced---possibly increasing
    voter confidence. 
  \item When a predetermined expected sampling rate is used, 
  the labor required can be estimated in advance, assuming
    escalation is not required.  With appropriate parameter choices,
    escalation can be avoided except in unusually close races, or when
    the reported outcome is wrong.  This helps
    election officials plan.
  \item If the sampling rate is selected after the reported
    margin is known, officials can choose a rate that makes
    escalation unlikely unless the reported electoral outcome is incorrect.
  \item The sampling approach is conceptually easy to grasp: toss a coin for
    each ballot.  The audit stops when the sample shows a sufficiently large margin for every
    winner over every loser, where ``sufficiently large'' depends on the sample size.
  \item The approach may have security advantages, since waiting longer
    to audit would leave more opportunity for the paper ballots to be
    compromised or misplaced.  Workers will need to handle the ballot papers
    in any case to move them from the ballot boxes into long-term storage.    
\end{itemize}

Officials selecting an auditing method should weigh these advantages
against some potential downsides of our approach, particularly when
applied in polling places on election night.  Poll workers are already
very busy, and they may be too tired at the end of the night to
conduct the sampling procedure or to do it accurately.  When audits
are conducted in parallel at local polling places, it is impossible
for an individual observer to witness all the simultaneous steps.
Moreover, estimating the sample size before margins are known makes it
likely that workers will end up sampling more (or fewer) ballots than
necessary to achieve the risk limit.  While sampling too little can be
overcome with escalation, the desire to avoid escalation may make
officials err on the side of caution and sample more than predicted to
be necessary, further reducing expected efficiency.

%\TODO{disadvantages: poll-workers may be too tired to do it, or to do it accurately; sample size might
%be too small or too large; hard for a single observer to see the audit in more than one polling place}
%\noindent \TODO{Still TODO (Philip):}
%
%\begin{itemize}
%    \item This may obviate the need to worry about stratification: different precincts take random samples independently, we get a random sample at the county level.
%\end{itemize}

\subsection{Previous Work}
Bernoulli sampling is a special case of Poisson sampling, where sampling units
are selected independently, but not necessarily with equal probability.
\citet{aslamEtal08} propose a Poisson sampling method in which the probability
of selecting a given unit is related to a bound on the error that unit could
hide.  
Their method is not an RLA: it is designed to have a large chance of detecting at least
one error if the outcome is incorrect, rather than to limit the risk of
certifying an incorrect outcome \emph{per se}.

\subsection{Stratified Audits}

Independent Bernoulli samples from different populations using the same rate 
still yields a Bernoulli sample of the overall population, so the math presented here can be used without modification
to audit contests that cross jurisdictional boundaries.
Bernoulli samples from different strata using different rates
can be combined using SUITE \citep{ottoboni2018risk}, which can be applied to
stratum-wise $P$-values from any method, including BBP.
(This requires minor modifications to the $P$-value calculations, to test arbitrary hypotheses
about the margin in each stratum rather than to test for ties; the derivations in
\cite{ottoboni2018risk} apply, \emph{mutatis mutandis}.)
If some ballots are tabulated using technology that makes a more efficient auditing approach
possible, such as a ballot-level comparison audit, it may be advantageous to 
stratify the ballots into groups, sample using Bernoulli sampling in some and a different method in
others, and use SUITE to combine the results into an overall RLA.

\section{Conclusion}

We presented a new ballot-polling RLA
based on Bernoulli sampling, relying on Wald's sequential probability ratio test.  
The new method performs similarly to the BRAVO ballot-polling audit but has several logistical
advantages, including that it can be parallelized and conducted on
election night, which may reduce cost and increase security.  
The method easily incorporates VBM and provisionally cast ballots,
and may eliminate the need for stratification in many circumstances.
Bernoulli ballot-polling with just a 1\% sampling rate
would have sufficed to confirm the 2016 U.S.\ Presidential
election results in the vast majority of states, if the reported results were correct.  
The practical benefits and conceptual simplicity of Bernoulli
ballot-polling may make it simpler to conduct risk-limiting audits in real elections.

% References: 1 page (can only fit 30-some citations, argh!)

%acknowledgements
{\footnotesize
\bibliographystyle{plainnat}
\bibliography{paper}}
%\nocite{*}
\appendix
\end{document}